\documentclass[twocolumn,english,aps,prb]{revtex4}
\usepackage[T1]{fontenc}
\usepackage[latin9]{inputenc}
\usepackage{color}
\definecolor{note_fontcolor}{rgb}{0.335938, 0, 1}
\usepackage{amsmath}
\usepackage{amssymb}
\usepackage{graphicx}

\makeatletter

\@ifundefined{textcolor}{}
{%
 \definecolor{BLACK}{gray}{0}
 \definecolor{WHITE}{gray}{1}
 \definecolor{RED}{rgb}{1,0,0}
 \definecolor{GREEN}{rgb}{0,1,0}
 \definecolor{BLUE}{rgb}{0,0,1}
 \definecolor{CYAN}{cmyk}{1,0,0,0}
 \definecolor{MAGENTA}{cmyk}{0,1,0,0}
 \definecolor{YELLOW}{cmyk}{0,0,1,0}
}

\makeatother

\usepackage{babel}
\begin{document}
\title{Non-ergodic extended phase of the Quantum Random Energy model}
\author{Lara Faoro$^{1,2}$, Mikhail V. Feigel'man$^{3,4}$ \& Lev Ioffe$^{1,2,5}$}
\affiliation{Laboratoire de Physique Theorique et Hautes Energies, Sorbonne Universite
and CNRS, France}
\affiliation{Department of Physics, University of Wisconsin, Madison, USA.}
\affiliation{L. D. Landau Institute for Theoretical Physics, Chernogolovka, 142432,
Moscow region, Russia}
\affiliation{Skolkovo Institute of Science and Technology, Moscow 143026, Russia}
\affiliation{Condensed Matter Physics Laboratory, National Research University
\textquotedbl Higher School of Economics\textquotedbl , Moscow,
Russia}
\begin{abstract}
The concept of non-ergodicity in quantum many body systems can be
discussed in the context of the wave functions of the many body system
or as a property of the dynamical observables, such as time-dependent
spin correlators. In the former approach the non-ergodic delocalized
states is defined as the one in which the wave functions occupy a
volume that scales as a non-trivial power of the full phase space.
In this work we study the simplest spin glass model and find that
in the delocalized non-ergodic regime the spin-spin correlators decay
with the characteristic time that scales as non-trivial power of the
full Hilbert space volume. The long time limit of this correlator
also scales as a power of the full Hilbert space volume. 
We identify this phase with the glass phase whilst the many body
localized phase corresponds to a 'hyperglass' in which dynamics is
practically absent. We discuss the implications of these finding to
quantum information problems. 
\end{abstract}
\maketitle
The ergodicity hypothesis states that the dynamic averaging is equivalent
to the ensemble averaging with statistical weight.\citep{Boltzmann1964}
Ergodicity is a common assumption in equilibrium statistical mechanics.
However, as it was first shown empirically by Kauzman in 1948,\citep{Kauzmann1948}
the ergodicity hypothesis fails in conventional glasses below the
vitrification transition. At temperatures below vitrification the
glass is locked into one of many metastable states, so the entropy
corresponding to the number of these states does not contribute to
the measurable quantities when the system is studied at reasonable
times scales. The appearance of the configurational entropy is a distinguishing
feature of the glass state and it is firmly established for many classical
glass models. Similar phenomenology has been shown in quantum glasses
with significant coupling to the environment: as the quantum dynamics
is reduced, the system is locked into one of the many metastable states.\citep{Cugliandolo2013}
As a result, for both classical glasses and dissipative quantum glasses,
it is believed that the glassy state is not completely frozen at all
non-zero temperatures and retains some amount of entropy. 

Dynamical properties of quantum glasses decoupled from environment
are much less understood. Because quantum glasses can be viewed as
disordered many body systems, at low temperature they exhibit many
body localization\citep{Basko2006}, that is the many body equivalent
of Anderson localization\citep{Anderson1958}. In the localized phase
their entropy is zero. However, it is not clear whether the quantum
glass transition is equivalent to the many body localization, in which
the system becomes completely frozen, or it leads to an intermediate
phase characterized by non-zero configurational and non-zero dynamically
accessible entropies similarly to classical and dissipative quantum
glasses. Since mathematically glass models are equivalent to optimization
problems, the answer to this question turns out to be relevant for
quantum computation where it translates into the estimate of the efficiency
of quantum algorithms. 

The dynamics of a many body system can be viewed as a particle hopping
on a graph of states in the Hilbert space.\citep{Altshuler1997} Recently,
a number of works reported the evidence for the existence of non-ergodic
delocalized phases for simplified models formulated directly in the
Hilbert space, such as disordered random regular graphs\citep{Altshuler2014,Altshuler2016,Kravtsov2018a,Khaymovich2018}
and Rosenzweig-Porter (RP)\citep{Kravtsov2015,Kravtsov2018b,Biroli2016}
models. Evidence for the non-ergodic extended (NEE) phase has been
also shown numerically in disordered Josephson junction chains.\citep{Pino2017}
In all these works, the non-ergodicity was defined as the property
of the eigenstates of the Hamiltonian, namely it was shown that the
effective volume occupied by these states scales as $\mathcal{N}^{D}$
where $0<D<1$ and $\mathcal{N}$ is the full volume of the Hilbert
space. Ergodicity corresponds to $D=1.$ However, the relation between
the non-ergodicity defined as the property of the eigenstates and
the one defined in glass physics remained unclear. 

\begin{figure}
\includegraphics[width=0.5\columnwidth]{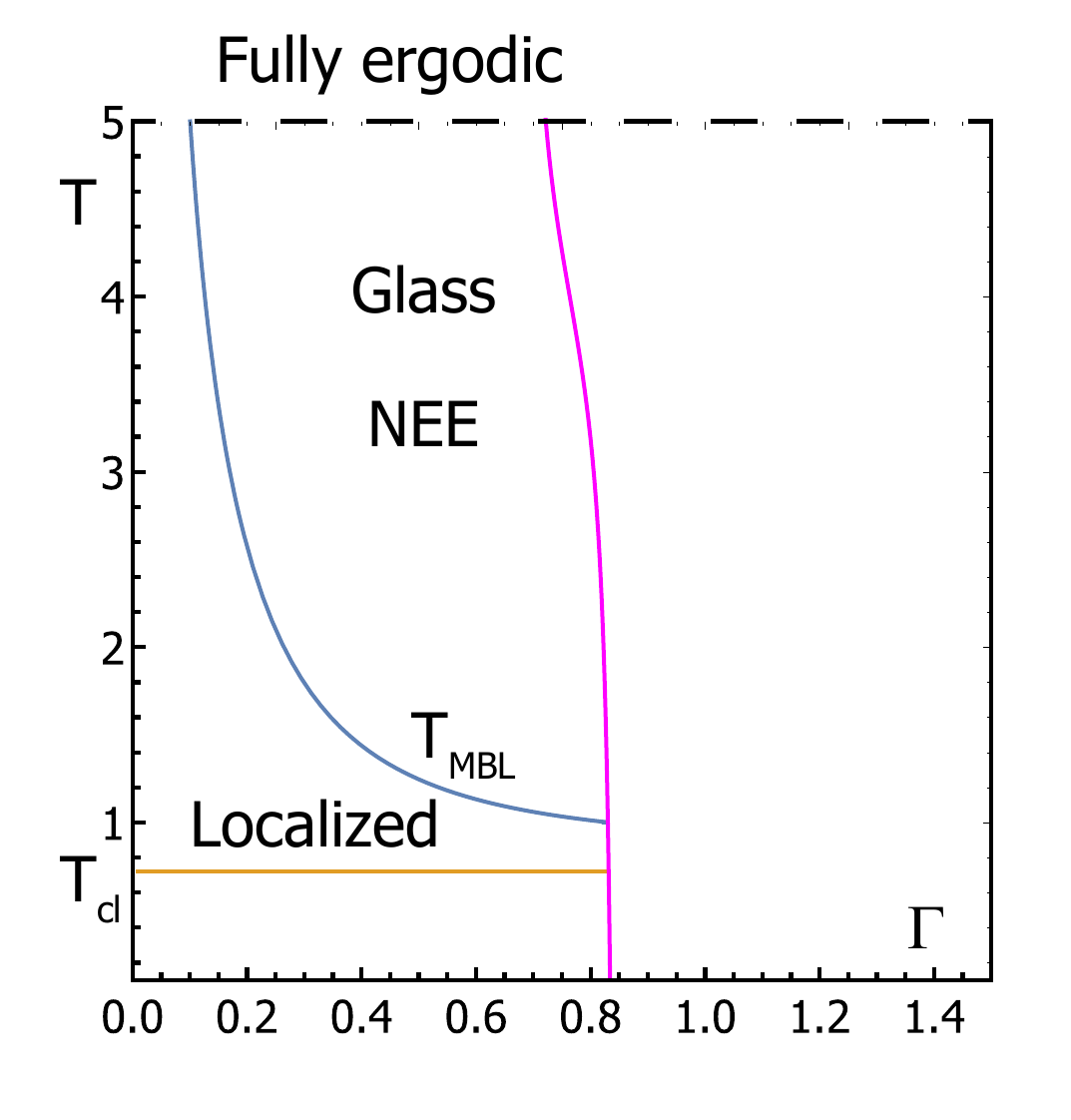}\includegraphics[width=0.5\columnwidth]{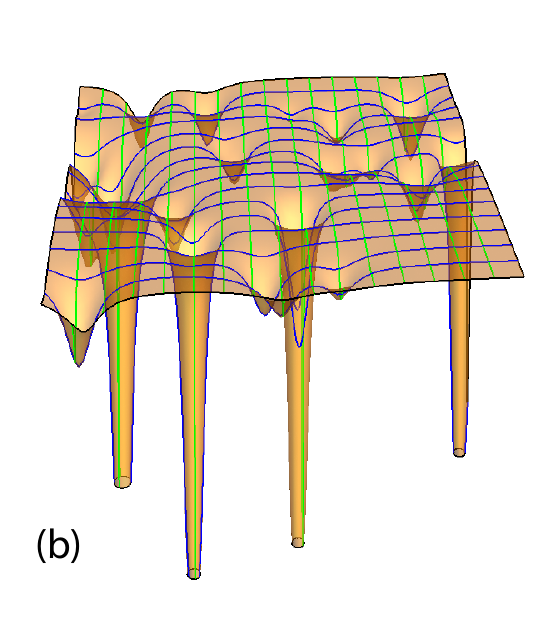}

\caption{(a) Phase diagram of qREM. Physically, the NEE phase corresponds to
the glassy dynamics whilst localized phase corresponds to a completely
frozen state, e.g. a hyperglass, see text. (b) Cartoon of 'golf course'
potential energy landscape that shows deep uncorrelated minima with
very small attraction basins. \label{fig:Phase-diagram} }
\end{figure}
{} In this work we fill this gap. Specifically, we study the simplest
quantum glass model that can be viewed as a simplest classical glass
in a trasnverse field. We show that the quantum dynamics of the low
energy states can be mapped to the one of the RP model. We focus on
the regime of small transverse fieds and find three distinct dynamical
phases: fully localized one in which the glass is completely frozen
(a hyperglass), non-ergodic extended one in which the spin dynamics
is slow but does not cease completely (a glass) and fully ergodic
paramagnetic state at high temperatures, see Fig. \ref{fig:Phase-diagram}a.
In addition, the full thermodynamic average displays the low temperature
transition from the glass to paramagnetic phases, this transition
was discussed in paper\citep{GoldSchmidt1990}. We also find that
the dynamical spin-spin correlator in the intermediate glassy phase
of this model displays the same scaling behavior as the survival probability
in the Hilbert space of the RP model studied in papers \citep{Kravtsov2018b,Biroli2016}.
Namely, both the dynamic glass order parameter defined by $q=\lim_{t\rightarrow\infty}\left\langle S(0)S(t)\right\rangle $
and the survival probability $R(t)=|\langle\Psi(t)|\Psi(0)\rangle|^{2}$
scale as $q\sim R(\infty)\sim\mathcal{N}^{-D}$. 

The fact that in a spin glass the dynamical order parameter and the
relaxation time scale as a power of the phase volume $\mathcal{N}$
implies the exponential scaling of these quantities with the number
of spins. This is not surprising for low energy states that are separated
from each other by large barriers and distances. In fact, similar
scaling was reported for the classical dynamics of the p-spin glass
in \citep{Lopatin2007}. 

We note that the problem of quantum dynamics in REM model was considered
previously in \citep{Laumann2014,Laumann2016}. This work correctly
notices that this model exhibits many body localization transition.
However, it incorrectly identifies the phase above MBL transition
with the quantum paramagnet. In fact, the dynamics in this phase involves
times that scale with the system size. It also uses forward propagation
approximation to obtain the the position of MBL line, which cannot
be justified in this problem. These mistakes are repeated in the subsequent
work \citep{Laumann2017} on p-spin model.

\subsubsection*{\textbf{Model and mapping to RP model.}}

The model we study is the quantum analogue of the Random Energy Model
(REM) introduced for classical glasses in a seminal work\citep{Derrida1980}.
The Hamiltonian of the qREM reads:
\begin{align}
H & =V\left(\{\sigma_{i}^{z}\}\right)-\Gamma\sum_{i=1}^{n}\sigma_{i}^{x}=H_{\text{REM}}+H_{\text{D}}\label{eq:H}\\
P(V) & =\frac{1}{\sqrt{\pi n}}\exp\left(-\frac{V^{2}}{n}\right)\label{eq:P(v)}
\end{align}
 The degrees of freedom in this model are $n$ spins. As in the REM,
$V\left(\{\sigma_{i}^{z}\}\right)$ is a function that takes $2^{n}$
different values for the $2^{n}$ configurations of the $n$ spins
in z-basis, $\{\sigma_{i}^{z}\}$. In the following we refer to these
spin configurations as spin words. The random potential values are
taken randomly from a Gaussian distribution $P(V)$ of zero mean and
variance $n/2$. Despite its simplicity $H_{\text{REM}}$ displays
many features of the glasses, such as the transition from the paramagnetic
state at high temperatures and the low temperature glass state in
which the partition function is dominated by the lowest energy state.
The transverse field $\Gamma$ is responsible for the dynamics of
the qREM, we consider only $\Gamma<1$ in this work. The thermodynamical
properties of the qREM are also well known\citep{GoldSchmidt1990,Kurchan2008,Laumann2016}:
at low transverse field, it displays the same transition as the classical
model between the paramagnetic and the glass phases. The qREM can
be mapped into an Anderson Model on $n$-dimensional hypercube, where
each spin word $\left\{ \sigma_{i}^{z}\right\} $ determines a site
with associated onsite energy $V(\{\sigma_{i}^{z}\})$. These potential
energies are completely uncorrelated in the full $\mathcal{N}=2^{n}$
dimensional space. The hopping between nearest neighbor sites is due
to the driver Hamiltonian $H_{D}=-\Gamma\sum_{i}\sigma_{i}^{x}$.

The distinguishing feature of the qREM given in (\ref{eq:H} and \ref{eq:P(v)})
is the absence of correlations in energy of the states separated by
just one spin flip. For the low energy states in the tail of the distribution
function $P(V)$ it implies that one spin flip takes the spin state
away from the low energy subspace. Qualitatively it corresponds to
the energy landscape that consists of rare narrow minima, similar
to a golf course with narrow deep holes, such as shown in Fig. \ref{fig:Phase-diagram}b.
Although unusual for ordinary spin glasses that always display strong
correlations between the energies of the states separated by one spin
flip, such energy landscape appears in the Number Partitioning Problem
that is equivalent to the REM model in the local energy subspace.\citep{Franz2004}
Because the quantum dynamics that starts from one low energy state
and leads to another can be viewed as a physical analogue of a quantum
search algorithm in a completely structureless problem \citep{Smelyanskiy2018a},
we expect that the golf course landscape will generally appear in
all problems that are equivalent to unstructured searches. 

The low temperature behavior of the qREM (\ref{eq:H},\ref{eq:P(v)})
is qualitatively different for small and large $\Gamma$: at small
$\Gamma$ the lowest energy states are due to the rare spin configurations
for which the potential $V$ is anomalously low, whilst at large $\Gamma$
they correspond to the spins polarized in $x$ direction. In this
work we shall focus on the regime of small $\Gamma\lesssim1$ for
which one expects the glassy behavior. We notice that the quantum
glass model (\ref{eq:H},\ref{eq:P(v)}) allows many modifications
relevant for the search algorithms; for instance Refs. \citep{Smelyanskiy2018a,Smelyanskiy2018b}
discussed the one in which $V=0$ for the majority of the configurations
whilst the remaining others are distributed in the narrow band energy
around $E_{0}\sim n$.

We focus on low temperature regime in which one expects a glassy behavior.
At these temperatures the behavior is controlled by the low energy
states. A distinguishing feature of the qREM is the presence of two
types of low energy states: the states originating from the low energy
configurations with anomalously small $V(\left\{ \sigma_{i}^{z}\right\} )$
and the states with large polarization in $x$-direction. It is convenient
to discuss them separately. In the absence of $\Gamma$, the qREM
Hamiltonian reduces to $V(\{\sigma_{i}^{z}\})=\sum_{i}^{2^{n}}V_{i}|z_{i}\rangle\langle z_{i}|$,
where we denote with $|z_{i}\rangle$ the spin states corresponding
to the spin words $\left\{ \sigma_{i}^{z}\right\} $. The spacing
between the levels at the energy per spin $\epsilon=E/n$ is
\begin{equation}
\delta=\sqrt{\frac{\pi}{n}}\exp\left(\epsilon^{2}-\ln2\right)n\label{eq:delta}
\end{equation}
At $\left|\epsilon\right|>\epsilon_{c}=\sqrt{\text{ln 2}}$ the spacing
becomes exponentially large because the spin states at these energies
are very rare. In a typical sample there are no states at energies
$\epsilon<\epsilon_{c}$. The temperature at which the partition function
is dominated by the lowest state with the energy $\epsilon\approx\epsilon_{c}$
corresponds to the glass transition in the classical model: $T_{c}=-1/(2\epsilon_{c})$.

The spectrum of the driver Hamiltonian can be written as: $H_{\text{D}}=-\Gamma\sum_{k=1}^{2^{n}}m_{x_{k}}|x_{k}\rangle\langle x_{k}|$,
where $m_{x_{k}}$denotes the polarization of the state $|x_{k}\rangle$.
$H_{\text{D}}$ has eigenvalues $\epsilon_{m}=-\Gamma m$ where $m=-n+2k$
with integer $k\in(0,n)$. Each discrete level has degeneracy $M(\epsilon_{m})=\left(\begin{array}{c}
n\\
k
\end{array}\right)\approx\text{exp[ln\ensuremath{2}-\ensuremath{m^{2}/2)]n} \ensuremath{\approx}}\exp[\ln2-\epsilon_{m}^{2}/(2\Gamma{}^{2})]n$. Comparing $M(\epsilon)$ with the density of spin states polarized
in the $z-$direction, $\delta^{-1}$ given in (\ref{eq:delta}),
we see that for $\Gamma<1/\sqrt{2}$ the states with $\epsilon\ll1$
are dominated by the classical ones in the limit $n\rightarrow\infty$.
The spectrum of polariazed states is bound by $\epsilon=\Gamma$,
so at very low energies the classical states dominate for $\Gamma<\ln^{1/2}2$. 

At temperatures, $T\gtrsim T_{c}$ the partition function is controlled
by the spin states with energies around $\epsilon=-1/(2T).$ At very
low temperature, the spin states are very far from each other, the
amplitude of quantum tunneling between them is much smaller than their
level spacing, so the quantum states remain fully localized . We term
this phase, in which the system remains completely frozen in the low
energy classical spin configurations, hyperglass. At higher energies,
$\epsilon>\epsilon_{A}$ Anderson delocalization happens. The transition
to the delocalized phase as well as the properties of this phase can
be analyzed by mapping the spin problem (\ref{eq:H} and \ref{eq:P(v)})
to the effective quantum problem of tunneling of low energy spin states
caused by the driver Hamiltonian $H_{D}$. The resulting effective
problem turns out to be equivalent to the RP model. 

The tunneling between spin words is due to the driver terms, $\Gamma\sigma_{i}^{x}$,
in the Hamiltonian. Because the density of the low energy spin states
is small, the tunneling between them appear only in high order of
the perturbation theory in $\Gamma$. We define the distance, $d$
between spin words as the minimal number of spin flips needed to get
from one to another. Because the number of the spin words grows exponentially
fast with the distance from a given state, the dynamics is dominated
by the spin states far away. Indeed, in the leading order of the perturbation
theory the amplitude, $H_{ab}=\left\langle z_{a}|H|z_{b}\right\rangle $
to tunnel the distance $d$ between two spin words corresponding to
the spin states $|z_{a}>$ and $|z_{b}>$ with energies $E\approx n\epsilon$
is $H_{ab}\sim(\Gamma/n\epsilon)^{d}d!$ while the number of spin
configurations at this distance increases as $\mathcal{B}(d)\sim n^{d}/d!$.
The condition to find a resonance at distance $d$, $\mathcal{H}_{ab}\mathcal{B}(d)\sim(\Gamma/\epsilon)^{d}\sim1$
is satisfied first at large $d$ indicating that large jumps are most
relevant. At large distances the number of spin words that one can
attain after $d$ spin flips is $P_{d}=\left(\begin{array}{c}
n\\
d
\end{array}\right)\approx\exp\left(n\ln2-(d-n/2)^{2}/2n\right)$. It has a sharp maximum at $d=n/2$. In the Methods section we show
that this dependence is faster than the decrease of the tunneling
amplitude with distance, so one can assume that a dominant tunneling
process between low energy states occurs due to jumps by distance
$d_{\text{typ}}\approx n/2$ . 

The tunneling between distant spin words can be evaluated by computing
such amplitude using only the driver part of the Hamiltonian, i.e.
neglecting the effect of the disorder potential on the tunneling.
Qualitatively, it means that when computing the tunneling between
deep holes in the golf course, Fig. \ref{fig:Phase-diagram}b we neglect
the effect of the small potential modulation between the deep holes.
To justify this assumption we note that the spectrum of the low energy
states polarized in $x$-direction is only weakly affected by the
disorder term. Indeed, the degeneracy of the $x$-polarized states
is lifted by the $V(\left\{ \sigma_{i}^{z}\right\} )$ term in the
Hamiltonian. To estimate this broadening we compute the effective
potential projected onto polarized states, $|x\rangle,|x'\rangle$
with the same polarization, $m$: $\text{\ensuremath{V_{xx'}=\langle x|V(\{\sigma_{i}^{z}\})|x'\rangle}}$.
The average value of this potential is zero, $\overline{V_{xx'}}=0$.
The matrix element between two states polarized in $z$ and $x$-directions
is $\left\langle z|x\right\rangle ^{2}=2^{-n}$, so
\begin{equation}
\overline{V_{xx'}^{2}}=n2^{-n-1}.\label{eq:V_xy}
\end{equation}
The random potential (\ref{eq:V_xy}) results in a level width that
remains much smaller than the distance between levels with different
polarizations:
\[
\Delta E=\left[\overline{V_{xx'}^{2}}M(\epsilon)\right]^{1/2}\sim\exp\left(-\frac{\epsilon_{m}^{2}}{4\Gamma^{2}}n\right)\ll\Gamma/n
\]
We conclude that the effect of the $V(\left\{ \sigma_{z}^{i}\right\} )$
term on the polarized states is small for the states at low energies. 

Crudely, one may estimate the amplitude of the tunneling by distance
$d_{\text{typ}}\approx n/2$ by noticing that a typical tunneling
process between two spin states, $\left|z_{a}\right\rangle $ and
$\left|z_{b}\right\rangle $ at low energies is due to the transitions
to highly delocalized states, $\left|\alpha\right\rangle $ in the
center of the band. The matrix element of this transition is $\left\langle z_{a}\left|\Gamma\sigma^{x}\right|\alpha\right\rangle \sim\left\langle z_{b}\left|\Gamma\sigma^{x}\right|\alpha\right\rangle \sim\Gamma2^{-n/2}$.
Thus, the contribution of a single delocalized state to the transition
between $\left|z_{a}\right\rangle $ and $\left|z_{b}\right\rangle $
is $(\Gamma^{2}/\epsilon n)2^{-n}$. The transition amplitudes have
random signs, so the summation over possible intermediate states gives
amplitude $\left|H_{ab}\right|\sim2^{-n/2}$. This estimate neglects
the orthogonality of the wave functions of the band center which leads
to the suppression of the amplitude for large $\epsilon$. The actual
computation that takes into account only the driving terms in the
Hamiltonian gives (see below for more details):

\begin{equation}
[(H_{ab})^{2}]_{typ}\approx e^{-n[\text{ln}2+\phi(\epsilon/\Gamma)]};\label{eq:H_ab}
\end{equation}
where the function

\begin{equation}
\phi(x)=\frac{1}{2}\ln(1-x^{2})+\frac{x}{2}\ln\frac{1+x}{1-x}\label{eq:phi}
\end{equation}
interpolates between $x^{2}/2$ at small $x$ and $\ln2$ at $x=1$.
At small $x$ the orthogonality of the wave functions in the center
of the band becomes irrelevant, so $\phi(x)\rightarrow0$ and one
reproduces the simple estimate above. 

The dominance of the tunneling to the most abundant spin words at
a given energy implies that the low energy sector can be mapped into
the RP model characterized by a ${\cal M}\times{\cal M}$ matrix Hamiltonian
with independent identically distributed fluctuating matrix elements
between the sites, $H_{mn},$ such that $\overline{(H_{n\neq m})^{2}}\sim{\cal M}^{-\gamma}$
and diagonal matrix elements (bare energies) $\overline{H_{nm}}=0$,
$\overline{(H_{nn})^{2}}=1$. At $\gamma>1$ the off-diagonal matrix
elements result in a hybridization of ${\cal M}^{D}$ states with
$D=2-\gamma<1$. For this hybridization only the states that are close
in energy, $E_{a}-E_{b}\sim H_{n\neq m}\ll1$ are relevant, which
implies that the model can be characterized by the typical distance
between adjacent bare energies, $\delta\sim\mathcal{M}^{-1}$, instead
of the the total number of states. The resulting model is defined
by $\overline{(H_{n\neq m})^{2}}\sim\delta^{\gamma}$. We expect it
to be equivalent to RP model for $\gamma>1$. 

At energy $\epsilon$ the distance between the energies of the adjacent
spin words scales as $\delta\sim\exp(-(\ln2-\epsilon^{2})n)$ so the
low energy states at this energies are equivalent to RP model with 

\begin{equation}
\gamma=\frac{\ln2+\phi(\epsilon/\Gamma)}{\ln2-\epsilon^{2}}\label{eq:gamma}
\end{equation}

The mapping of the low energy sector to the RP model allows us to
establish the presence of two dynamical transitions in the qREM model
in addition to the static transition at $\epsilon^{2}=\ln2$. Indeed,
the RP model has three distinct phases\citep{Kravtsov2015,Kravtsov2018b}:
for $0<\gamma<1$ the systems is ergodic and has fractal dimension
$D=1,$ for $1\leq\gamma\leq2$ the system is in a non ergodic extended
phase and has corresponding fractal dimension $D=2-\gamma$ and finally
for $\gamma>2$ the system is localized corresponding to fractal dimension
$D=0$. In qREM the transition to the ergodic phase at $\gamma=1$
occurs at $\epsilon\rightarrow0$, that is at energies $E\ll n$.
We shall not discuss this transition in this work. At all fixed $\epsilon>0$
the qREM is non-ergodic and it becomes completely localized at $\epsilon<-\epsilon_{A}$
where $\epsilon_{A}$ is determined by the condition $\gamma=2$.
Note that the position of the MBL transition line is not given by
the formula $|\epsilon|\approx\Gamma$ proposed in Refs.~\citep{Laumann2014,Laumann2016}. 

In the non-ergodic regime the return probability, $R(t)=\overline{|\langle\Psi_{a}(0)|\Psi_{a}(t)\rangle|^{2}}$,
in the equivalent RP model can be found by arguing that off-diagonal
matrix elements result in the decay of the initial state with energy
$E_{a}$ into a bath of states with close energies. Such process is
irreversible, it is described by a simple exponential relaxation with
the relaxation rate 
\begin{equation}
1/\tau=\delta^{\gamma-1}\label{eq:1/tau_RP}
\end{equation}

At very long time the wave function becomes spread over all states
belonging to the same mini-band. As a result, the probability to find
the system in the initial state scales as $P\sim\delta^{2-\gamma}$
at long times. Thus one concludes that in RP model the return probability
is given by $R(t)\approx e^{-t/\tau}+R_{\infty}$. This conclusion
was verified by simulations\citep{Kravtsov2018b} and computations
\citep{Biroli2016}. Because the spin configurations at distance $d_{\text{typ}}\approx n/2$
are completely uncorrelated with each other, the same arguments can
be applied to the spin-spin correlator:
\begin{align}
\left\langle \sigma_{i}^{z}(t)\sigma_{i}^{z}(0)\right\rangle  & \approx\exp(-t/\tau)+q_{EA}\label{eq:S(t)S(0)}\\
1/\tau\sim & \exp(-\theta n)\label{eq:1/tau_REM}\\
q_{EA} & \sim\exp(-\eta n)\label{eq:q_EA}
\end{align}
 where 
\begin{equation}
\theta=\epsilon^{2}+\phi(\epsilon/\Gamma)\label{eq:theta}
\end{equation}
\begin{equation}
\eta=\ln2-2\epsilon^{2}-\phi(\epsilon/\Gamma)\label{eq:eta}
\end{equation}

\begin{figure}
\includegraphics[width=0.8\columnwidth]{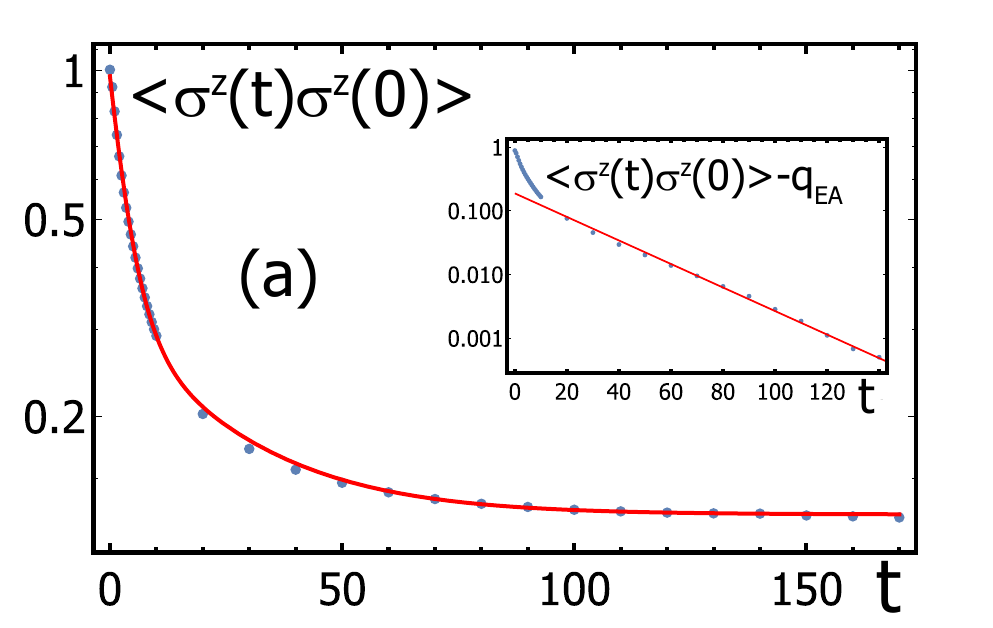}

\includegraphics[width=0.8\columnwidth]{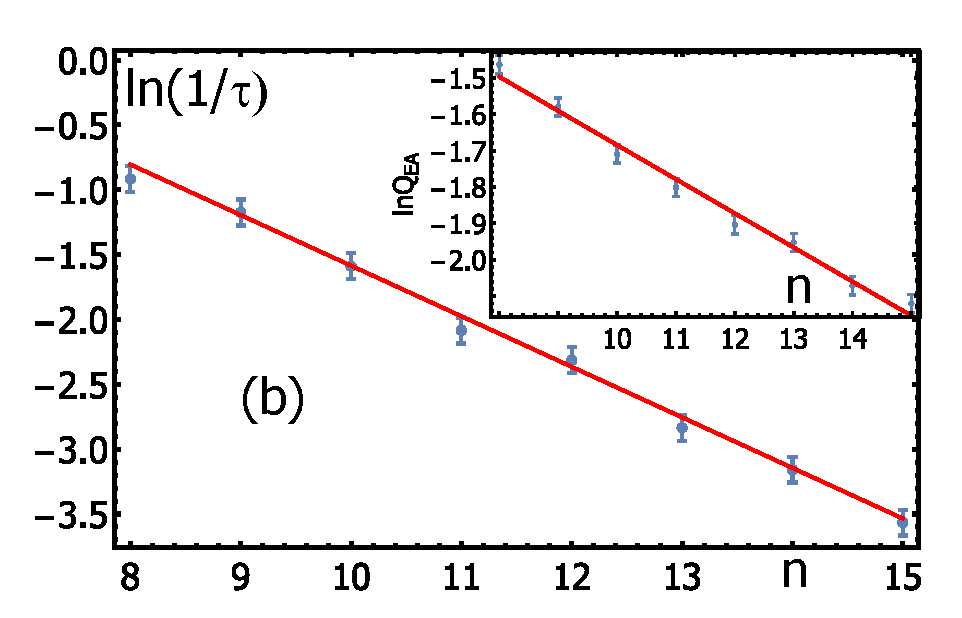}

\caption{(a) Spin-spin correlator in qREM model (\ref{eq:1/tau_REM}-\ref{eq:eta})
in non-ergodic delocalized regime corresponding to $\epsilon=-0.35$,
$\Gamma=0.5$ for $n=14$. The logarithmic plot of $\left\langle \sigma_{i}^{z}(t)\sigma_{i}^{z}(0)\right\rangle -q_{EA}$
shows that it follows exponential dependence (\ref{eq:S(t)S(0)})
over two orders of magnitude. (b) Size dependence of relaxation rate,
$1/\tau$ and $q_{EA}$ for $\epsilon=0.35$, $\Gamma=0.5$. \label{fig:Spin-spin-correlator}}
\end{figure}

In order to verify the predictions (\ref{eq:1/tau_REM}-\ref{eq:eta})
of the qualitative reasoning we have performed direct simulations
of the dynamics of the model (\ref{eq:H}\ref{eq:P(v)}) in the regime
of parameters where one expects to observe the anomalous dimension
$0<D<1$. The characteristic behavior of the spin-spin correlator
obtained in these simulations for $\Gamma=0.5$ and $T=1/(2\left|\epsilon\right|)$
with $\epsilon=-0.35$ is shown in Fig. \ref{fig:Spin-spin-correlator}a.
As expected it displays exponential decrease (\ref{eq:S(t)S(0)})
to a constant value. The relaxation rate follows the exponential dependence
(\ref{eq:1/tau_REM}) with the exponent $\theta\approx0.39$ expected
for these parameters. The dynamic spin glass order parameter, $q_{EA}$
also follows exponential size dependence (\ref{eq:q_EA}) with exponent
$\eta\approx0.1$ that is slightly smaller than the expected value
$\eta\approx0.17$. 

The quantum process that starts from the low energy state and leads
to another low energy state can be viewed as a solution of the computational
problem in which one searches for a state with the full quantum energy
that is close to the initial one. The results above imply that it
succeeds after time $\tau$ given by $\tau=\exp(\theta n)$. The search
leads to one of $\exp(\eta n)$ states, so the classical time to find
one of such states by brute force search scales as $\tau_{cl}=\exp((\ln2-\eta)n)$
multiplied by the time needed to evaluate the quantum energies. Clearly,
the quantum time is much shorter than the classical one in whole range
of delocalized states, reminiscent of the Grover search\citep{Grover1997}. 

\subsubsection*{\textbf{Conclusions}}

The qREM model defined by (\ref{eq:H}\ref{eq:P(v)}) can be viewed
as the simplest many body model that displays localization in its
Fock's space. We believe that the appearance of the intermediate non-ergodic
state is not an exotic feature but a typical behavior for many body
disordered models. 

The non-ergodic nature of the wave functions is fragile: broadening
of each level that remains non-zero in the thermodynamic limit destroys
it. This has important implication for the recent studies\citep{Sondhi2018}
that uses random unitary circuit as a toy model for chaotic many body
quantum systems. Because random circuit dynamics can be viewed as
continuous dynamics with external noise, it is very likely that one
cannot observe non-ergodicity in these studies. 

Empirically, the classical glasses can be divided into the ones with
highly correlated, funnel-like energy landscape and the ones with
weak correlations.\citep{Angell2008} We expect the mapping to RP
model to hold for the quantum version of the later at low energies
but not for the former. In contrast to the RP dynamics discussed here
that is characterized by a single relaxation time, the correlated
energy landscape display a continuous spectrum of the relaxation times.

The continuous spectra of relaxation times appears in many glass models
in which the barriers between local minima scale with $n$. In the
limit of $n\rightarrow\infty$ at any finite $t$ such glass is trapped
in the threshold state characterized by a broad spectrum of relaxation
times, the phenomena known as ageing. \citep{CuKu1993,Bouchaud1996,Cugliandolo2013}.
The dynamics discussed in this work occurs in the opposite limit,
at the time scales that are exponentially long in $n$. In this limit
the glass is able to explore lower energy states that can be often
viewed as deep uncorrelated minima. In particular, it has been shown
that classical dynamics of the p-spin model is dominated by the states
far in the configuration space separated by the large flat barrier.\citep{Lopatin2007}
It is very likely that quantum version of the p-spin model is equivalent
to RP. 

The relaxation of the a given low energy state can be viewed as a
quantum search process. As a result of this search one finds a state
which energy is close to the initial one. Because the energies of
the initial and final state include quantum corrections, it is not
straightforward to translate the results of this search into the algorithm
which results can be checked on a classical computer. We leave this
question for future work. We note that recent papers\citep{Laumann2018}
claiming the solution of this important problem ignore the crucial
requirement of the classical verification, constructing thereby, in
the words of Scott Aaronson a ``computer that simulates itself''.\citep{Aaronson2019}
Other important questions that remain to be addressed are the possibility
to implement the qREM Hamiltonian in superconducting circuits in order
to solve numerical problems such as NPP and the sensitivity of the
results to small coupling to the environment and dissipation. 

An interesting corollary of our work are the implications for the
quantum information scrambling in many body systems. Because of the
non-ergodicity, we expect the appearance of a wide regime in which
chaotic dynamics leads to incomplete information scrambling despite
the fact that all the regime of non ergodic delocalized states can
be viewed as chaotic, characterized by wave function spreading and
growth of von Neumann entropy. However, the information is far from
being spread uniformly over all allowed states.

{} A very recent work\citep{Laumann2018} on quantum optimization finds
an intermediate phase that it calls ``tunneling'' but does not realize
that it is the spin glass phase characterized by anomalously long
times, it also makes incorrect claim that in this phase $q_{EA}=0$
(i.e. $x=1/2$ in the notations of \citep{Laumann2018}) in contrast
with (\ref{eq:q_EA}). We note that $x=1/2$ is in the apparent contradiction
with the numerical data shown in Fig. 7 of the same work\citep{Laumann2018}
that clearly shows $x<1/2$. 

\textbf{Acknowledgments} We are grateful to Boris Altshuler, Kostyantyn
Kechedzhi, Vladimir Kravtsov and Vadim Smelyanskiy for useful discussions.
The work was partially supported by ARO grant ARO grant W911NF-13-1-0431. 

\subsection*{Details of the computation.}

\subsubsection*{\textbf{Matrix elements: analytic derivation and numerical results}}

Here we calculate the Green function $G_{ab}^{(0)}(E)$ that determines
the transition amplitude between the local (in the $z-$ basis) states
$a,b$ with energy $E$, separated by the distance $d_{ab}=\rho N$
on the hypercube. In this computation we neglect the presence of the
random potential $V_{a}$; it can be justified for the contribution
to $G_{ab}$ that comes from the $x-$ polarized states with extensive
$\sum_{i}\langle\sigma_{i}^{x}\rangle\propto n$ because these states
are weakly affected by the random potential.

We start from the Green function in the imaginary-time representation:
\begin{eqnarray}
G^{(0)}(\tau,\rho) & = & \prod_{i}(\cosh\Gamma\tau+\sigma_{i}^{x}\sinh\Gamma\tau)=\label{eq:G(tau,rho)}\\
 & = & \cosh^{N(1-\rho)}(\Gamma\tau)\sinh^{N\rho}(\Gamma\tau)\nonumber 
\end{eqnarray}
where we took into account that the product (\ref{eq:G(tau,rho)})
contains exactly $N\rho$ operators $\sigma_{i}^{x}$. Energy spectrum
of the kinetic part of the Hamiltonian is limited to the stripe $E\in(-N\Gamma,+N\Gamma)$.
For the energies $E=N\epsilon$ outside of this band (that is, $|\epsilon|>\Gamma$)
the Green function can be found from (\ref{eq:G(tau,rho)}) by the
Laplace transform and further saddle-point integration (using $N\gg1$
condition): 
\begin{equation}
G^{(0)}(\epsilon,\rho)\approx\exp\left[-NF\left(\frac{\epsilon}{\Gamma},\rho\right)\right]\label{eq:G(epsilon,rho)}
\end{equation}
where 
\begin{equation}
F(y,\rho)=y\tau^{*}-(1-\rho)\ln\cosh(\tau^{*})-\rho\ln\sinh(\tau^{*})\label{eq:F(x,rho)}
\end{equation}
and the saddle-point value of $\tau$, $\tau_{sp}=\tau^{*}/\Gamma$
is determined by 
\begin{equation}
y=(1-\rho)\tanh(\tau^{*})+\rho\coth(\tau^{*})\label{eq:y}
\end{equation}
The set of equations (\ref{eq:G(epsilon,rho)},\ref{eq:F(x,rho)},\ref{eq:y})
simplifies for the most relevant case of $\rho=\frac{1}{2}$, leading
to 
\begin{equation}
F\left(y,\frac{1}{2}\right)=y\tanh^{-1}\left(y-\sqrt{y^{2}-1}\right)+\frac{1}{4}\ln\left(y^{2}-1\right)+\frac{1}{2}\ln(2).\label{G5}
\end{equation}
In order to find the Green function at the energies inside the \textquotedbl conduction
band\textquotedbl , $|\epsilon|<\Gamma$, we employ analytic continuation
of (\ref{G5}) over $y$ into the range $|y|<1$, to obtain the result
(\ref{eq:H_ab},\ref{eq:phi}). 

The analytical computation of the transition amplitude neglects completely
the effect of the disorder on the tunneling. In order to check the
validity of this approximation, to check the mapping of qREM to RP
and establish the parameters of the RP model we performed a number
of numerical simulations of qREM model. 

First, we computed the time dependent spin-spin correlator and extracted
the exponents $\theta$ and $\eta$ that determine the relaxation
time, $\tau$, and the spin-glass order parameter. Comparing the numerical
results with analytical expectations we conclude that both $\tau$
and $q_{EA}$ display exponential dependencies on the system size
$n$ as expected (\ref{eq:1/tau_REM},\ref{eq:q_EA}) in the whole
non-ergodic phase. The exponents controlling these dependencies are
very close to the expected values at energies away from localization
transition. However, the difference between expected and observed
exponents become significant at low energies. In particular, the localization
transition occurs at significantly lower energies than expected analytically
indicating larger tunneling amplitudes than the ones given by (\ref{eq:H_ab},\ref{eq:phi}).
This enhancement of the tunneling amplitude can be viewed as renormalization
of the effective $\Gamma$ that increases the density of polarized
states at low energies. 

\begin{figure}
\includegraphics[width=0.8\columnwidth]{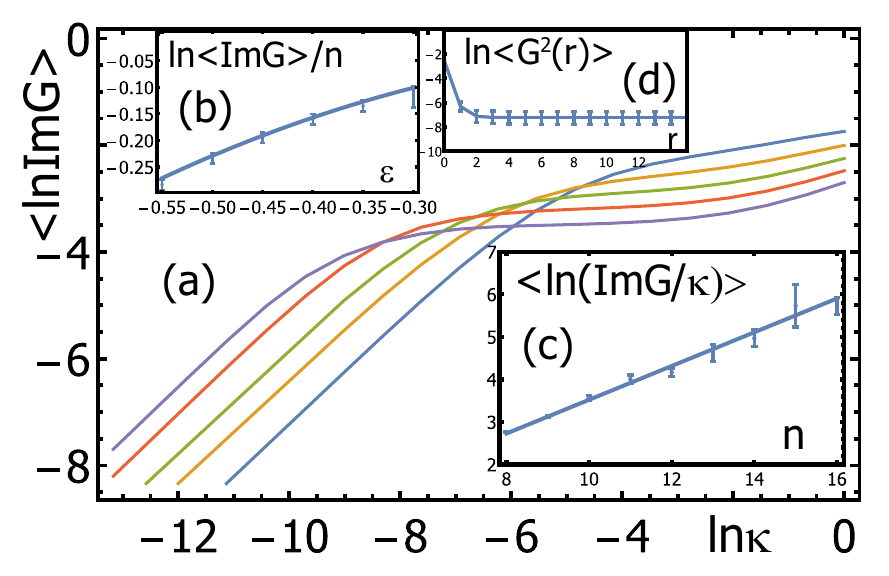}

\includegraphics[width=0.8\columnwidth]{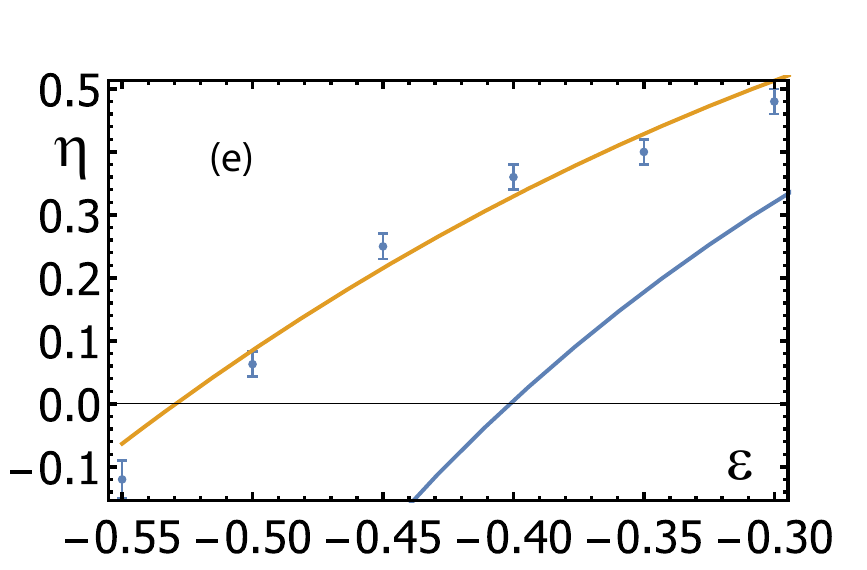}

\caption{(a) Typical $\Im\ensuremath{G(\epsilon,\kappa)}$ for $\epsilon=-0.35,$
$\Gamma=0.5$ as a function of $\kappa$ for $n=8,10,12,14,16$. (b)
Data for $\left\langle \Im\ensuremath{G(\epsilon,\kappa)}\right\rangle $
and their fit with $\ln\left\langle \Im\ensuremath{G(\epsilon,\kappa)}\right\rangle \approx-0.8n\epsilon^{2}.$
(c) The size dependence of the slope $\ln\left[\Im\ensuremath{G(\epsilon,\kappa)}/\kappa\right]$
for small $\kappa$ on the system size $n$. (d) Green function dependence
on the distance. (e) Value of $\eta$ determined from $d\ln\left[\Im\ensuremath{G(\epsilon,\kappa)}/\kappa\right]/dn$
and the analytical result corresponding to renormalized $\tilde{\Gamma}=0.7$
as discussed in the text. The lower line shows the analytical result
expected for $\Gamma=0.5$. \label{fig:Typical-ImG}}
\end{figure}
Second, we have computed the diagonal part of $\Im\ensuremath{G(\epsilon,\kappa)}$
defined by $G(\epsilon,\kappa)=(\epsilon n-H-i\kappa)^{-1}$. The
average value of this quantity gives the density of states; it is
close to the expected $\ln\left\langle \Im\text{Tr}G\right\rangle =-n\epsilon^{2}$
as shown in Fig. \ref{fig:Typical-ImG}. In contrast, the\emph{ typical}
value of $\Im\ensuremath{G(\epsilon,\kappa)}$ is controlled by the
matrix element that couples a given site to the resonance site at
energy $E=-n\epsilon$. At small $\kappa<\delta$, the dominant contribution
comes from the spin word that is closest in energy to $E$, so $\Im\ensuremath{G(\epsilon,\kappa)}_{\text{typ}}=\kappa[(H_{ab})^{2}]_{\text{typ}}/\delta^{2}\sim\kappa\exp(-n\eta)$,
where $\delta$ is the level spacing at energy $E$. At larger $\kappa>\delta$
the dominant contribution comes from many levels and typical value
of $\Im\ensuremath{G(\epsilon,\kappa)}$ saturates at the value given
by the Fermi golden rule $\Im\ensuremath{G(\epsilon,\kappa)}_{\text{typ}}\sim\exp(-n\theta)$.
Exactly this behavior is observed numerically, see Fig. \ref{fig:Typical-ImG}.
However, similar to spin-spin correlator, the precise values of the
exponents $\theta$ and $\eta$ determined from these simulations
differ somewhat from their analytical values. In particular, the density
of states becomes $\nu(\epsilon)=\exp(-c\epsilon^{2}n)$ with $c=0.8$
instead of $c=1.0$ while faster than expected $n$-dependence of
$\Im\ensuremath{G(\epsilon,\kappa)}_{\text{typ}}$ indicates renormalization
of the effective $\Gamma.$ For instance, for $\Gamma=0.5$ the value
of $\eta$ fits well the equation $\eta=\ln2-2c\epsilon^{2}-\phi(\epsilon/\tilde{\Gamma})$
with $\tilde{\Gamma}\approx0.7$. For high energies, $\epsilon\gtrsim-0.35$
the values of $\theta$ extracted from the $\Im G$ plateau agree
very well with the relation $\theta=\ln2-c\epsilon^{2}-\eta$. Unfortunately
it is not possible to check this relation for the whole range of energies
because the plateau in $\Im G(\eta)$ is not well defined at low energies
for the available sizes $n\lesssim16$. Finally, the off-diagonal
Green function shows very little dependence on the distance for $d\gg1$
(see Fig. \ref{fig:Typical-ImG}d) that justifies the assumption that
the tunneling processes are dominated by hops to large distances $d\sim n/2$. 

\begin{figure}
\includegraphics[width=0.8\columnwidth]{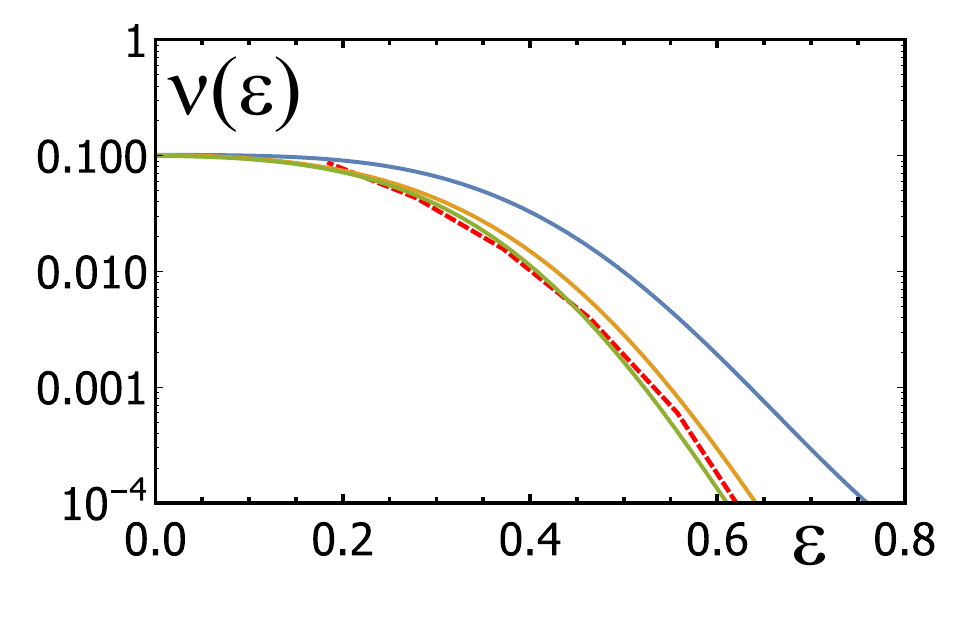}

\caption{Density of states with large magnetization in $x$-direction. Two
lower full curves show $\nu_{\perp}(\epsilon)$ for $\mu=0.25$ and
$\mu=0.5$ at $\Gamma=0.50$ The dashed line shows the density of
polarized states expected for $\tilde{\Gamma}=0.70$. Finally, the
upper curve shows the full density of states $\nu(\epsilon)$. \label{fig:Density-of-states}}
\end{figure}

Third, we have checked directly the renormalization of $\Gamma$ by
studying the density of polarized states defined by $\nu_{\perp}(\epsilon)=\exp(-\mu n\epsilon/\tilde{\Gamma})\text{Tr}\exp\left(\mu\sum\sigma^{x}\right)\Im G(\epsilon)$.
The additional factor $\exp\left(\mu\sum\sigma^{x}\right)$ gives
extra weight $\exp(\mu m)$ to the states with magnetization $m$
that selects spin states with large magnetizations along $x$-direction.
For instance at $\epsilon=\Gamma$ the contribution of the fully polarized
state gets additional factor $\exp(\mu n)$ that overweights the density
of classical states $\exp\left(\ln2-\Gamma^{2}\right)n$ for $\mu>\ln2-\Gamma^{2}$.
The numerical simulation shows that the density of polarized states
defined in this way is indeed $\mu$ independent and coincides with
the one expected for the model with renormalized $\tilde{\Gamma}$
as shown in Fig. \ref{fig:Density-of-states}. 

These results make us believe that the presence of intermediate energy
states affects the tunneling between the low energy states, which
can be described as an increase in the value of $\Gamma\rightarrow\tilde{\Gamma}$.
The same hybridization also increases the apparent density of states
leading to $c<1$. In order to exclude the finite size effects we
performed a separate study of a modified model in which $V(\{\sigma_{i}^{z}\})=0$
for most spin words and randomly distributed around $-\epsilon_{0}$
for others and verified that the renormalization of $\Gamma$ is absent
in this model as expected.\citep{Smelyanskiy2018b} 

\bibliography{MBLandNonequilibrium}

\end{document}